\documentclass[aps,prb,twocolumn,longbibliography,showpacs]{revtex4-1}

\usepackage{graphicx}
\usepackage{dcolumn}
\usepackage{bm}
\usepackage[version=3]{mhchem}
\usepackage{color}

\begin{document}

\title{Polaron formation, native defects, and electronic conduction in metal tungstates}
\author{Khang Hoang}
\email[E-mail: ]{khang.hoang@ndsu.edu}
\affiliation{Department of Physics \& Center for Computationally Assisted Science and Technology, North Dakota State University, Fargo, North Dakota 58108, USA.}

\date{\today}

\begin{abstract}

Iron tungstate (\ce{FeWO4}) and manganese tungstate (\ce{MnWO4}) belong to a family of wolframite-type materials that has applications in various areas, including supercapacitors, batteries, and multiferroics. A detailed understanding of bulk properties and defect physics in these transition-metal tungstates has been lacking, however, impeding possible improvement of their functional properties. Here, we report a first-principles study of \ce{FeWO4} and \ce{MnWO4} using screened hybrid density-functional calculations. We find that in both compounds the electronic structure near the band edges are predominantly the highly localized transition-metal $d$ states, which allows for the formation of both hole polarons at the Fe (Mn) sites and electron polarons at the W sites. The dominant native point defects in \ce{FeWO4} (\ce{MnWO4}) under realistic synthesis conditions are, however, the hole polarons at the Fe (Mn) sites and negatively charged Fe (Mn) vacancies. The presence of low-energy and highly mobile polarons provides explanation for the good p-type conductivity observed in experiments and the ability of the materials to store energy via a pseudocapacitive mechanism.    

\end{abstract}


\maketitle


\section{Introduction}\label{sec;intro}

\begin{figure}
\vspace{0.2cm}
\centering
\includegraphics[width=0.8\linewidth]{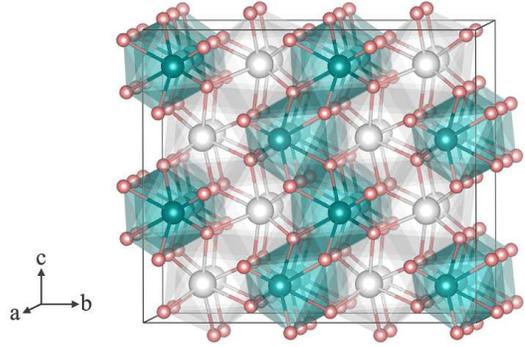}
\vspace{-0.1cm}
\caption{(color online) Crystal structure of \ce{FeWO4}. Large (gray) spheres are Fe, medium (blue) spheres are W, and small (red) spheres are O. The structure of \ce{MnWO4} is similar.}
\label{fig;struct}
\end{figure}

Wolframite-type iron tungstate (\ce{FeWO4}), see Fig.~\ref{fig;struct}, has recently been investigated as a new pseudocapacitive electrode material for high volumetric energy density supercapacitors operated in an aqueous electrolyte.\cite{Bretesche2015} The superior performance of nanocrystalline \ce{FeWO4} is thought to originate from Fe$^{3+}$/Fe$^{2+}$ fast redox reactions at the surface. The material has also been considered for lithium-ion battery anodes.\cite{Shim2010} It was found back in the 1980s that \ce{FeWO4} is a ``p-type semiconductor''; the electronic conduction was suggested to be governed by polaron hopping processes with the activation energy in the range 0.15$-$0.32 eV.\cite{Sieber1982,Sieber1983,Schmidbauer1991} The electrical transport properties of the Mn analog of \ce{FeWO4}, i.e., manganese tungstate (\ce{MnWO4}), had also been studied with the activation energy reported to be higher, in the range 0.53$-$0.65 eV.\cite{Bharati1982,Dissanayake1989} More recently, the material has been of interest for supercapacitors\cite{Li2015} and multiferroics.\cite{Heyer2006,Choi2010,Ruiz-Fuertes2012,Patureau2016} Effects of the Mn:W non-stoichiometry on the ferroelectric transition have also been investigated,\cite{Yu2014} though the nature of the chemical disorder is still unknown. 

Computational studies of \ce{FeWO4} and \ce{MnWO4} in particular and transition-metal tungstates in general have been scarce. Rajagopal et al.,\cite{Rajagopal2010} for example, investigated the electronic structure of \ce{FeWO4} and \ce{CoWO4} using density-functional theory (DFT) and the standard generalized-gradient approximation for the exchange-correlation functional. However, it is well known that the method cannot describe properly the physics of complex transition-metal oxides. Ruiz-Fuertes et al.\cite{Ruiz-Fuertes2012} studied the electronic structure of \ce{MnWO4} (and several other metal tungstates) using the DFT$+U$ extension; however, the on-site Coulomb interaction $U$ term was applied only on the Mn $d$ states. As a result, the calculated band gap is severely underestimated compared to the reported experimental values.\cite{Ruiz-Fuertes2012} More importantly, first-principles studies of defects in the metal tungstates have completely been lacking, yet in order to understand the above-mentioned experimental observations one needs to have a detailed understanding of the defect physics. 

We herein report a study of bulk properties, polaron formation and migration, and native defect landscapes in \ce{FeWO4} and \ce{MnWO4}, using a hybrid DFT/Hartree-Fock method. We find that while the electronic structure allows for the formation of both hole and electron polarons, the dominant defects in the tungstates under realistic synthesis conditions are hole polarons and negatively charged Fe (Mn) vacancies. In light of the results, we discuss the electronic conduction in the materials and comment on the observed redox pseudocapacitance.

\section{Methodology}\label{sec;method}

Our calculations employ the Heyd-Scuseria-Ernzerhof (HSE06) screened hybrid functional,\cite{heyd:8207} the projector augmented wave (PAW) method,\cite{PAW1} and a plane-wave basis set, as implemented in the Vienna {\it Ab Initio} Simulation Package (\textsc{vasp}).\cite{VASP1,VASP2,VASP3} We use the standard PAW potentials in the \textsc{vasp} database which treat Fe $3d^74s^1$, Mn $3d^64s^1$, W $6s^25d^4$, and O $2s^22p^4$ explicitly as valence electrons and the rest as core electrons. The Hartree-Fock mixing parameter and the screening length are set to the standard values of 25\% and 10 {\AA}, respectively; the plane-wave basis-set cutoff is set to 500 eV. The calculations for bulk \ce{FeWO4} and \ce{MnWO4} (two formula units per unit cell) are carried out using a 3$\times$3$\times$3 \textbf{k}-point mesh; for other bulk phases, \textbf{k}-point meshes are chosen such that the \textbf{k}-point density stays almost the same (the smallest allowed spacing between \textbf{k}-points is fixed at 0.5 {\AA}$^{-1}$). A denser, 5$\times$4$\times$5, \textbf{k}-point mesh is used in calculations to obtain electronic densities of states and dielectric constants. Defects are modelled using 2$\times$2$\times$2 (96-atom) supercells. Integrations over the Brillouin zone in defect calculations is carried out using the $\Gamma$ point (Our computational tests using a denser, 2$\times$2$\times$2, Monkhorst-Pack \textbf{k}-point mesh gives a formation-energy difference of about 5 meV). In all calculations, structural relaxations are performed with HSE06 and the force threshold is chosen to be 0.01 eV/{\AA}; spin polarization is included.

The formation energy of a defect or defect complex X in effective charge state $q$ is defined as
\begin{align}\label{eq;eform}
E^f({\mathrm{X}}^q)&=&E_{\mathrm{tot}}({\mathrm{X}}^q)-E_{\mathrm{tot}}({\mathrm{bulk}}) -\sum_{i}{n_i\mu_i} \\ %
\nonumber &&+~q(E_{\mathrm{v}}+\mu_{e})+ \Delta^q ,
\end{align}
where $E_{\mathrm{tot}}(\mathrm{X}^{q})$ and $E_{\mathrm{tot}}(\mathrm{bulk})$ are, respectively, the total energies of a supercell containing X and of an equivalent supercell of the perfect bulk material. $\mu_{i}$ is the atomic chemical potential of species $i$ that have been added to ($n_{i}$$>$0) or removed from ($n_{i}$$<$0) the supercell to form the defect. $\mu_{e}$ is the electronic chemical potential, i.e., the Fermi level, referenced to the valence-band maximum (VBM) in the bulk ($E_v$). $\Delta^q$ is the correction term to align the electrostatic potentials of the bulk and defect supercells and to account for finite-size effects on the total energies of charged defects,\cite{walle:3851} determined following the approach of Freysoldt, Neugebauer, and Van de Walle.~\cite{Freysoldt} Defects with lower formation energies will form more easily and occur in higher concentrations.

The atomic chemical potentials $\mu_{i}$ are subject to thermodynamic constraints. For example, the stability of the host compound \ce{FeWO4} requires
\begin{equation} 
\mu_{\mathrm{Fe}} + \mu_{\mathrm{W}} + 4\mu_{\mathrm{O}} = \Delta H(\mathrm{FeWO}_4), 
\end{equation}
where $\Delta H$ is the formation enthalpy. Similarly, the stability condition for \ce{MnWO4} is
\begin{equation} 
\mu_{\mathrm{Mn}} + \mu_{\mathrm{W}} + 4\mu_{\mathrm{O}} = \Delta H(\mathrm{MnWO}_4). 
\end{equation}
Further constraints involve the requirement that the host compound is stable against all competing Fe$-$W$-$O (or Mn$-$W$-$O) phases; see Sec.~\ref{subsec;chempot}. We note that, in our current work, the zero reference state of $\mu_{\mathrm{Fe}}$ (or $\mu_{\mathrm{Mn}}$) and $\mu_{\mathrm{W}}$ is the total energy per atom of the respective bulk metals, whereas the reference of $\mu_{\mathrm{O}}$ is chosen to be half of the total energy of an isolated \ce{O2} molecule at 0 K.\cite{Hoang2015PRA}  

\section{Results and discussion}\label{sec;results}

\subsection{Bulk properties}\label{subsec;bulk}

The tungstates \ce{FeWO4} and \ce{MnWO4} are isostructural compounds, crystallizing in a monoclinic structure (space group $P2/c$); see Fig.~\ref{fig;struct}. The lattice parameters obtained in our HSE06 calculations are $a=4.761$ {\AA}, $b=5.693$ {\AA}, $c=4.990$ {\AA}, and $\beta=90.16^\circ$ for \ce{FeWO4}; $a=4.831$ {\AA}, $b=5.806$ {\AA}, $c=4.986$ {\AA}, and $\beta=91.18^\circ$ for \ce{MnWO4}. These values are all in excellent agreement with the experimental ones: $a=4.753$ {\AA}, $b=5.720$ {\AA}, $c=4.968$ {\AA}, and $\beta=90.08^\circ$ for \ce{FeWO4},\cite{Escobar1971} and $a=4.830$ {\AA}, $b=5.7603$ {\AA}, $c=4.994$ {\AA}, and $\beta=91.14^\circ$ for \ce{MnWO4}.\cite{Macavei1993} In \ce{FeWO4}, iron is stable as high-spin Fe$^{2+}$ with a calculated magnetic moment of 3.69 $\mu_{\rm B}$, consistent with experiments.\cite{Sieber1982} Manganese in \ce{MnWO4} is stable as high-spin Mn$^{2+}$ with a calculated magnetic moment of 4.57 $\mu_{\rm B}$. In a simple ionic model, \ce{FeWO4} (\ce{MnWO4}) can be regarded as consisting of Fe$^{2+}$ (Mn$^{2+}$), W$^{6+}$, and O$^{2-}$. 

The electronic contribution to the static dielectric constant of \ce{FeWO4} (\ce{MnWO4}) is about 4.86 (4.57) in HSE06, based on the real part of the dielectric function $\epsilon_{1}(\omega)$ for $\omega\rightarrow0$. The ionic contribution is calculated using density-functional perturbation theory, \cite{Wu2005,dielectricmethod} within the generalized-gradient approximation.\cite{GGA} The total dielectric constants are 18.95 and 17.39 for \ce{FeWO4} and \ce{MnWO4}, respectively, in excellent agreement with the reported experimental values of 19.6 and 19.7.\cite{rieck2013}

In addition to the ferromagnetic (FM) structure, we also explore two antiferromagnetic (AF) configurations of \ce{FeWO4} (\ce{MnWO4}): one (AF1) with parallel spins within the Fe (Mn) zigzag chains along the $c$-axis but with adjacent chains coupled antiferromagnetically, and the other (AF2) with antiparallel spins within the Fe (Mn) chains. All three configurations are investigated using 2$\times$1$\times$1 supercells and a 2$\times$3$\times$3 \textbf{k}-point mesh. We find that in both compounds, the AF and FM spin configurations are almost degenerate in energy. Specifically, AF2 in \ce{FeWO4} is higher in energy than FM and AF1 by only 6 meV per formula unit (f.u.); in \ce{MnWO4}, AF1 and AF2 are lower than FM by 10 meV/f.u. Our results are thus in contrast to those for \ce{FeWO4} reported by Almeida et al.\cite{almeida2012} where the total-energy difference between the AF and FM spin configurations was shown to be much larger.    

\begin{figure*}
\vspace{0.2cm}
\centering
\includegraphics[width=0.82\linewidth]{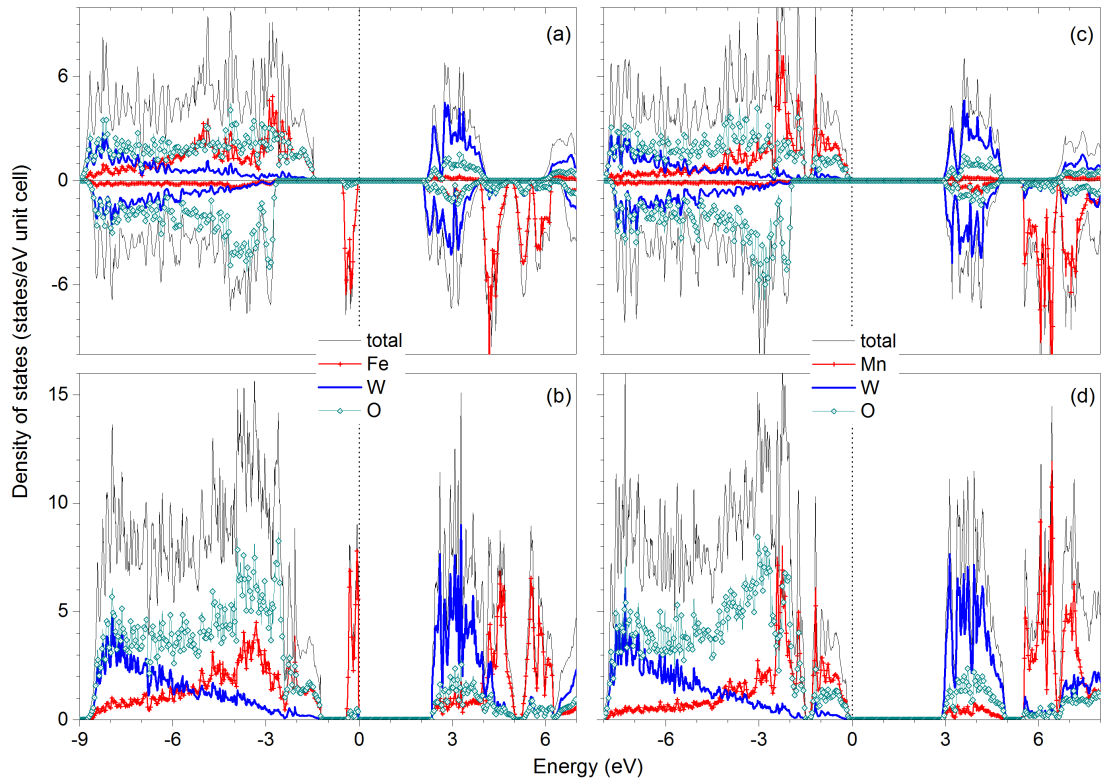}
\vspace{-0.1cm}
\caption{(color online) Total and projected density of states of \ce{FeWO4} (a) without and (b) with spin-orbit coupling and \ce{MnWO4} (c) without and (d) with spin-orbit coupling. The zero of energy is set to the highest occupied states. }
\label{fig;dos}
\end{figure*}

Figure \ref{fig;dos} shows the electronic structure of \ce{FeWO4} and \ce{MnWO4} without and with spin-orbit coupling (SOC), all in the FM spin configuration. We find that in both compounds the VBM is predominantly the highly localized Fe (Mn) $3d$ states, whereas the conduction-band minimum (CBM) is predominantly the empty W $5d$ states. The calculated band gap is 2.08 eV (without SOC) or 2.33 eV (with SOC) for \ce{FeWO4}; 2.92 eV (without SOC) or 2.88 eV (with SOC) for \ce{MnWO4}. For comparison, the reported experimental band gap is 2.0 eV for \ce{FeWO4},\cite{Ejima2006} or in the range 2.37$-$3.0 eV for \ce{MnWO4}.\cite{Ruiz-Fuertes2012,Choi2010,Parhi2008,Ejima2006} Given the highly localized  transition-metal $d$ states at the VBM and CBM, the formation of hole and electron polarons is expected in both \ce{FeWO4} and \ce{MnWO4} (which is indeed the case, as will be discussed in Sec.~\ref{subsec;defect}). We note that, except for the small differences in the calculated band gap values, SOC does not change the nature of the electronic structure near the band edges; see Fig.~\ref{fig;dos}. SOC can therefore be excluded in our supercell defect calculations to save on computing time since its inclusion does not change the physics of what we are presenting.  

\subsection{Chemical potentials}\label{subsec;chempot}

\begin{figure*}
\vspace{0.2cm}
\centering
\includegraphics[width=0.85\linewidth]{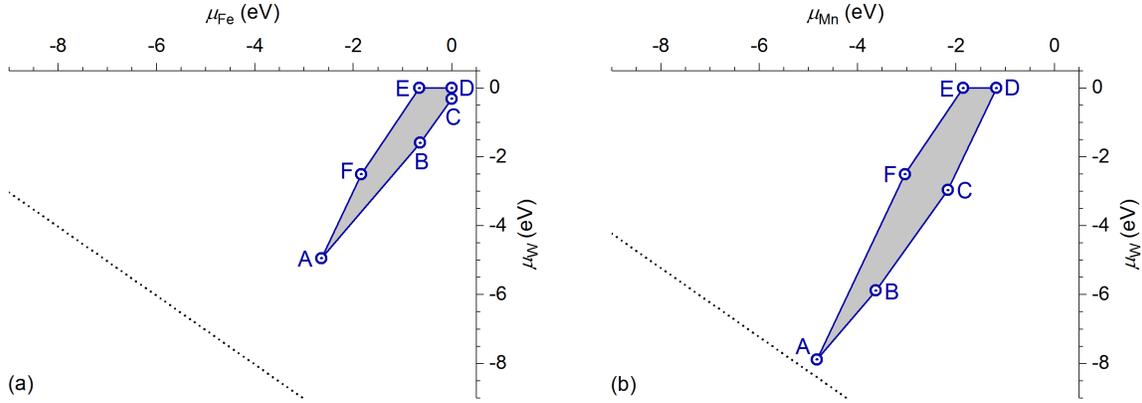}
\caption{(color online) Chemical-potential diagrams for (a) \ce{FeWO4} and (b) \ce{MnWO4}. Only the Fe$-$W$-$O (or Mn$-$W$-$O) phases that define the stability region of \ce{FeWO4} (\ce{MnWO4}), here shown as a shaded polygon, are included. In the case of \ce{FeWO4}, these phases are Fe$_2$O$_3$ (along AB), Fe$_3$O$_4$ (BC), Fe (CD), W (DE), W$_{18}$O$_{49}$ (EF), and WO$_3$ (FA); as for \ce{MnWO4}, the phases are Mn$_2$O$_3$ (along AB), Mn$_3$O$_4$ (BC), Mn$_3$WO$_6$ (CD), W (DE), W$_{18}$O$_{49}$ (EF), and WO$_3$ (FA). The O$_2$ phase at 0 K (the dotted line, where the oxygen chemical potential $\mu_{\rm O} = 0$ eV) is also included for reference.}
\label{fig;chempot}
\end{figure*}

\begin{table}
\caption{Calculated formation enthalpies at 0 K, in eV per formula unit. The experimental standard enthalpies of formation, $\Delta H^0$(298.15 K), are also included.}\label{tbl;enthalpies}
\begin{center}
\begin{ruledtabular}
\begin{tabular}{llrr}
Compound & Crystal structure &This work & $\Delta H^0$(298.15 K) \\
\colrule
\ce{FeWO4} &Monoclinic& $-$12.03 & $-$11.97 (Ref.\cite{Dean}) \\
\ce{Fe2WO6} &Orthorhombic& $-$16.72 &\\
\ce{FeO} &Tetragonal& $-$2.60 & $-$2.82 (Ref.\cite{chase}) \\
\ce{Fe2O3} &Monoclinic& $-$8.63 & $-$8.56 (Ref.\cite{chase}) \\
\ce{Fe3O4} &Cubic& $-$11.72 & $-$11.62 (Ref.\cite{chase}) \\
\ce{MnWO4} &Monoclinic& $-$13.23 & $-$13.53 (Ref.\cite{Martins2014}) \\
\ce{Mn3WO6} &Trigonal& $-$21.62 & \\
\ce{MnO} &Cubic & $-$4.12 & $-$3.96 (Ref.~\cite{Knacke1991}) \\
\ce{MnO2} &Tetragonal & $-$4.86 & $-$5.41 (Ref.~\cite{Knacke1991}) \\
\ce{Mn2O3} &Orthorhombic & $-$10.05 & $-$9.94 (Ref.~\cite{Knacke1991})\\
\ce{Mn3O4}&Tetragonal & $-$14.61 & $-$14.37 (Ref.~\cite{Knacke1991}) \\
\ce{WO2} &Tetragonal  & $-$5.11 & $-$6.11 (Ref.\cite{chase}) \\
\ce{WO3} &Orthorhombic& $-$8.27 & $-$8.74 (Ref.\cite{chase}) \\
\ce{W18O49} &Monoclinic & $-$139.24 & $-$145.80 (Ref.\cite{chase}) \\
\end{tabular}
\end{ruledtabular}
\end{center}
\end{table}

Figure \ref{fig;chempot} shows the calculated chemical-potential diagrams for \ce{FeWO4} and \ce{MnWO4}, constructed by exploring all possible phases in the Fe$-$W$-$O and Mn$-$W$-$O phase spaces (see Table \ref{tbl;enthalpies}), respectively. The shaded polygon is where the host compound is thermodynamically stable. In principle, $\mu_{\rm Fe}$ ($\mu_{\rm Mn}$), $\mu_{\rm W}$, and $\mu_{\rm O}$ can have any values within this region. However, we will be interested in a smaller part of the stability region in which the atomic chemical potentials represent conditions that are close to actual experimental conditions. For example, \ce{FeWO4} is often prepared from a solid-state reaction of \ce{Fe2O3} (or FeO) and \ce{WO3} at 900$^\circ$C.\cite{Sieber1982,Schmidbauer1991} These conditions are expected to be within the region spanning from the $\mu_{\rm O}=-1.11$ eV level (at point A where there is an equilibrium between \ce{FeWO4}, \ce{Fe2O3}, and \ce{WO3}) to the $\mu_{\rm O}=-1.92$ eV level (that contains point F where \ce{FeWO4}, \ce{WO3}, and \ce{W18O49} are in equilibrium); see Fig.~\ref{fig;chempot}(a). We note that $\mu_{\rm O}=-1.11$ and $-1.92$ eV correspond to the oxygen chemical potential in air at about 685 and 1275$^\circ$C, respectively.\cite{Reuter2001} \ce{MnWO4}, on the other hand, can be prepared from a solid-state reaction of MnO and \ce{WO3} at 600$^\circ$C in air.\cite{Yu2014} The material is also synthesized using other methods at lower temperatures.\cite{Dissanayake1989,Patureau2016} The synthesis conditions of \ce{MnWO4} are thus expected to be approximately within the region spanning from the $\mu_{\rm O}=-0.13$ eV level (at point A where \ce{MnWO4}, \ce{Mn2O3}, and \ce{WO3} are in equilibrium) to the $\mu_{\rm O}=-1.92$ eV level (that contains point F where \ce{MnWO4}, \ce{WO3}, and \ce{W18O49} are in equilibrium); see Fig.~\ref{fig;chempot}(b).    

\subsection{Defect landscapes}\label{subsec;defect}    

\begin{figure}
\vspace{0.2cm}
\centering
\includegraphics[width=0.95\linewidth]{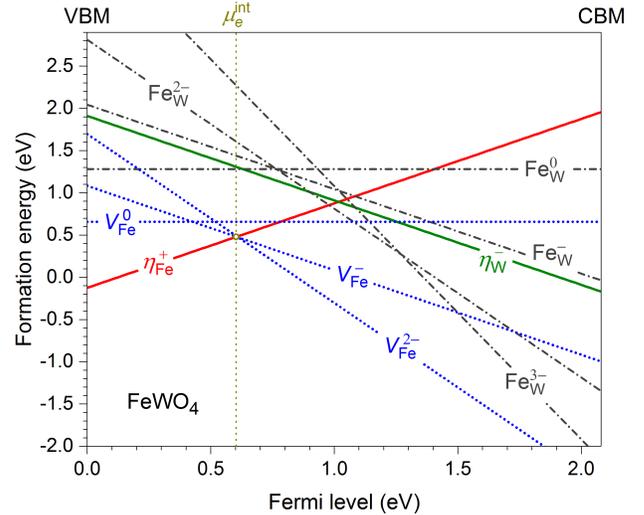}
\vspace{-0.1cm}
\caption{(color online) Formation energies of relevant point defects in \ce{FeWO4} obtained under conditions at point A in Fig.~\ref{fig;chempot}(a), plotted as a function of Fermi level from the VBM to the CBM. The slope indicates the charge state: positively (negatively) charged defects have positive (negative) slopes. $\mu_e^{\rm int}$, marked by the vertical dotted line, is the position of the Fermi level determined by charge neutrality condition.}
\label{fig;fe;fewo4}
\end{figure}

\begin{figure}
\vspace{0.2cm}
\centering
\includegraphics[width=0.95\linewidth]{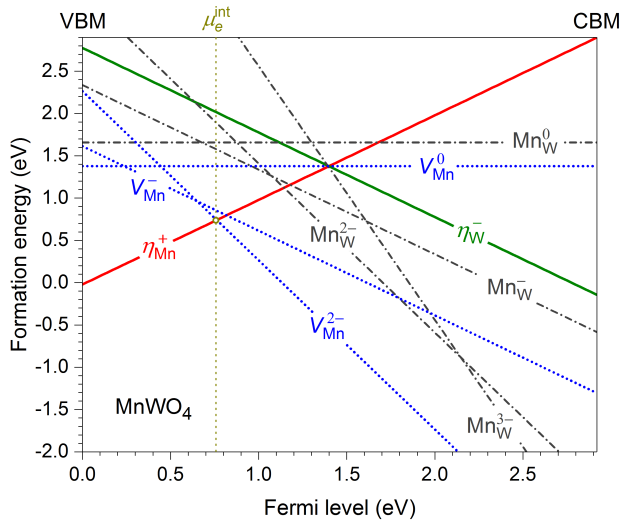}
\vspace{-0.1cm}
\caption{(color online) Formation energies of relevant point defects in \ce{MnWO4} obtained under conditions at point A in Fig.~\ref{fig;chempot}(b), plotted as a function of Fermi level from the VBM to the CBM. $\mu_e^{\rm int}$, marked by the vertical dotted line, is the position of the Fermi level determined by charge neutrality.}
\label{fig;fe;mnwo4}
\end{figure}

\begin{table*}
\caption{Calculated formation energies ($E^{f}$) and binding energies ($E_{b}$) of relevant point defects in the tungstates \ce{FeWO4} and \ce{MnWO4}, obtained at points A$-$F in Figs.~\ref{fig;chempot}(a) and \ref{fig;chempot}(b). The transition-metal ion associated with each elementary defect is listed in parentheses.}\label{tab:formenergy}
\begin{center}
\begin{ruledtabular}
\begin{tabular}{llcccccclr}
&&\multicolumn{6}{c}{$E^{f}$ (eV)}& \\
\cline{3-8}
&Defect&A&B&C&D&E&F&Constituents&$E_{b}$ (eV) \\
\colrule
\ce{FeWO4}&$\eta_{\rm Fe}^{+}$&0.48&0.89&0.89&0.89&0.89&0.75&(Fe$^{3+}$)\\ 
&$\eta_{\rm W}^{-}$ &1.30&0.89&0.89&0.89&0.89&1.03&(W$^{5+}$)\\ 
&Fe$_{\rm W}^{3-}$&2.26&2.37&3.01&3.32&3.98&3.07&(Fe$^{3+}$)\\  
&Fe$_{\rm W}^{2-}$&1.60&2.12&2.76&3.07&3.73&2.68&Fe$_{\rm W}^{3-} + \eta_{\rm Fe}^{+}$&1.14\\  
&Fe$_{\rm W}^{-}$&1.44&2.37&3.01&3.32&3.98&2.79&Fe$_{\rm W}^{3-} + 2\eta_{\rm Fe}^{+}$&1.79\\
&Fe$_{\rm W}^{0}$&1.28&2.62&3.26&3.57&4.23&2.91&Fe$_{\rm W}^{3-} + 3\eta_{\rm Fe}^{+}$&2.43\\
&W$_{\rm{Fe}}^{3+}$&3.72&3.61&2.97&2.66&2.00&2.90&(W$^{5+}$)\\   
&W$_{\rm{Fe}}^{2+}$&3.47&2.95&2.31&2.00&1.34&2.38&(W$^{4+}$)\\
&W$_{\rm{Fe}}^{+}$&3.84&2.91&2.27&1.96&1.29&2.48&(W$^{3+}$)\\ 
&W$_{\rm{Fe}}^{0}$&4.80&3.46&2.82&2.51&1.85&3.17&W$_{\rm{Fe}}^{+} + \eta_{\rm W}^{-}$&0.34\\
&$V_{\rm{Fe}}^{2-}$&0.48&1.67&2.31&2.31&1.65&0.75&\\   
&$V_{\rm{Fe}}^{-}$&0.48&2.08&2.72&2.72&2.05&1.02&$V_{\rm{Fe}}^{2-} + \eta_{\rm Fe}^{+}$&0.49\\ 
&$V_{\rm{Fe}}^{0}$&0.66&2.67&3.31&3.31&2.64&1.47&$V_{\rm{Fe}}^{2-} + 2\eta_{\rm Fe}^{+}$&0.79\\
&$V_{\rm{O}}^{2+}$&3.13&2.61&2.13&2.05&2.22&2.86&\\  
&$V_{\rm{O}}^{+}$&3.31&2.38&1.90&1.82&1.99&2.77&$V_{\rm{O}}^{2+} + \eta_{\rm W}^{-}$&1.12 \\ 
&$V_{\rm{O}}^{0}$&3.84&2.50&2.02&1.94&2.11&3.02&  \\
\ce{MnWO4}&$\eta_{\rm Mn}^{+}$&0.74&1.14&1.38&1.41&1.56&1.34&(Mn$^{3+}$)\\ 
&$\eta_{\rm W}^{-}$ &2.01&1.61&1.38&1.34&1.19&1.42&(W$^{5+}$)\\ 
&Mn$_{\rm W}^{3-}$&3.27&2.87&3.62&5.49&5.72&5.07&(Mn$^{3+}$)\\  
&Mn$_{\rm W}^{2-}$&1.89&1.89&2.87&4.78&5.16&4.28&(Mn$^{4+}$)\\  
&Mn$_{\rm W}^{-}$&1.57&1.97&3.20&5.13&5.67&4.56&Mn$_{\rm W}^{2-} + \eta_{\rm Mn}^{+}$&1.05\\  
&Mn$_{\rm W}^{0}$&1.66&2.46&3.92&5.89&6.57&5.25&Mn$_{\rm W}^{2-} + 2\eta_{\rm Mn}^{+}$&1.71\\  
&W$_{\rm{Mn}}^{3+}$&4.13&4.53&3.78&1.91&1.68&2.33&(W$^{5+}$)\\   
&W$_{\rm{Mn}}^{2+}$&4.46&4.46&3.48&1.57&1.19&2.07&(W$^{4+}$)\\
&W$_{\rm{Mn}}^{+}$&5.33&4.93&3.71&1.77&1.24&2.34&(W$^{3+}$)\\ 
&W$_{\rm{Mn}}^{0}$&6.98&6.18&4.72&2.75&2.07&3.39&W$_{\rm{Mn}}^{+} + \eta_{\rm W}^{-}$&0.36\\ 
&$V_{\rm{Mn}}^{2-}$&0.74&1.14&2.13&3.05&2.07&1.34&\\   
&$V_{\rm{Mn}}^{-}$&0.85&1.65&2.87&3.83&3.00&2.05&$V_{\rm{Mn}}^{2-} + \eta_{\rm Mn}^{+}$&0.63\\ 
&$V_{\rm{Mn}}^{0}$&1.38&2.58&4.04&5.03&4.34&3.17&$V_{\rm{Mn}}^{2-} + 2\eta_{\rm Mn}^{+}$&0.84\\
&$V_{\rm{O}}^{2+}$&2.89&2.89&2.26&1.34&1.82&2.29&\\  
&$V_{\rm{O}}^{+}$&3.72&3.32&2.46&1.51&1.83&2.53& $V_{\rm{O}}^{2+} + \eta_{\rm W}^{-}$&1.18\\ 
&$V_{\rm{O}}^{0}$&4.90&4.10&3.00&2.02&2.19&3.11& \\ 
\end{tabular}
\end{ruledtabular}
\end{center}
\end{table*}

Figures \ref{fig;fe;fewo4} and \ref{fig;fe;mnwo4} show the calculated formation energies of low-energy defects in \ce{FeWO4} and \ce{MnWO4}, obtained under conditions associated with point A in Figs.~\ref{fig;chempot}(a) and \ref{fig;chempot}(b), respectively. These defects include hole polarons associated with iron ($\eta^+_{\rm Fe}$, i.e., Fe$^{3+}$ at the Fe$^{2+}$ site) or manganese ($\eta^+_{\rm Mn}$, i.e., Mn$^{3+}$ at the Mn$^{2+}$ site), electron polarons ($\eta^-_{\rm W}$, i.e., W$^{5+}$ at the W$^{6+}$ site), and iron (Fe$_{\rm W}$) or manganese (Mn$_{\rm W}$) antisites and iron ($V_{\rm Fe}$) or manganese ($V_{\rm Mn}$) vacancies in different charge states. The results for oxygen vacancies ($V_{\rm O}$) and interstitials (O$_i$) and tungsten antisites (W$_{\rm Fe}$ and W$_{\rm Mn}$) are not included the figures as they have much higher formation energies. In the absence of electrically active impurities that can shift the Fermi-level position or when such impurities occur in much lower concentrations than charged native defects, the Fermi level is at $\mu_{e}^{\rm int}$, determined only by the native/intrinsic defects (Here, we also assume that free holes and electrons in the materials are negligible). With the chosen set of the atomic chemical potentials, the Fermi level of \ce{FeWO4} is at $\mu_{e}^{\rm int}=0.61$ eV, determined predominantly by the hole polaron $\eta^+_{\rm Fe}$ and the negatively charged iron vacancy $V_{\rm Fe}^{2-}$, as seen in Fig.~\ref{fig;fe;fewo4}; for \ce{MnWO4}, $\mu_{e}^{\rm int}$ is at 0.76 eV, determined predominantly by $\eta^+_{\rm Mn}$ and $V_{\rm Mn}^{2-}$, as seen in Fig.~\ref{fig;fe;mnwo4}. The position of the Fermi level $\mu_{e}^{\rm int}$ as well as the defect landscape change as one changes the atomic chemical potentials (which represent the experimental conditions); see Eq.~(\ref{eq;eform}). The evolution of the defect landscape is reflected in Table~\ref{tab:formenergy} where we report the formation energies at $\mu_{e}^{\rm int}$ of all relevant defects in \ce{FeWO4} and \ce{MnWO4}, obtained at different points in the chemical-potential diagrams in Figs.~\ref{fig;chempot}(a) and \ref{fig;chempot}(b).

Since the VBM of \ce{FeWO4} is predominantly the highly localized Fe $3d$ states [see Fig.~\ref{fig;dos}(a)], the removal of an electron from the supercell to create $\eta_{\rm Fe}^+$ results in a highly localized hole at one of the Fe sites, turning one Fe$^{2+}$ into a high-spin Fe$^{3+}$ ion with a calculated magnetic moment of 4.23 $\mu_{\rm B}$. The average Fe$^{3+}$$-$O bond length is 2.05 {\AA}, compared to 2.14 {\AA} of the Fe$-$O bonds in the bulk. The local lattice environment is thus distorted in the presence of the localized hole. The formation of $\eta_{\rm W}^-$ involves adding an electron to the supercell, resulting in a highly localized electron at one of the W sites, i.e., turning one W$^{6+}$ into a W$^{5+}$ ion with a magnetic moment of $-$0.71 $\mu_{\rm B}$. This can be understood from the electronic structure of \ce{FeWO4} in which the CBM is predominantly W $5d$ states [see Fig.~\ref{fig;dos}(a)]. The average W$^{5+}$$-$O bond length is 1.98 {\AA}, compared to 1.94 {\AA} of the W$-$O bonds in the bulk. Similarly, $\eta_{\rm Mn}^+$ in \ce{MnWO4} is high-spin Mn$^{3+}$ with a magnetic moment of 3.84 $\mu_{\rm B}$ plus local lattice distortion. The average Mn$^{3+}$$-$O bond length is 2.07 {\AA}, compared to 2.19 {\AA} of the Mn$-$O bonds in the bulk. $\eta_{\rm W}^-$ in \ce{MnWO4} is W$^{5+}$ plus local lattice distortion, similar to $\eta_{\rm W}^-$ in \ce{FeWO4}. In both compounds, the hole and electron polarons can be regarded as small polarons. The self-trapping energy ($E_{\rm ST}$), defined as the difference between the formation energy of the free hole or electron and that of the hole or electron polaron,\cite{Hoang2014} is calculated to be 0.28 eV for $\eta_{\rm Fe}^+$ and 0.17 eV for $\eta_{\rm W}^-$ in \ce{FeWO4}; $E_{\rm ST}$ = 0.28 eV for $\eta_{\rm Mn}^+$ and 0.12 eV for $\eta_{\rm W}^-$ in \ce{MnWO4}. The hole polarons thus have higher $E_{\rm ST}$, which is consistent with the fact that the local lattice environment of $\eta_{\rm Fe}^+$ ($\eta_{\rm Mn}^+$) is more distorted than that of the electron polarons $\eta_{\rm W}^-$. 

Regarding other defects, the transition-metal-related defect configurations are either elementary defects associated with the transition-metal ion in its stable charge states in \ce{FeWO4} and \ce{MnWO4} (Fe$^{2+,3+}$, Mn$^{2+...4+}$, or W$^{3+...6+}$) or defect complexes consisting of an elementary defect and hole/electron polaron(s). For example, $V_{\rm{Fe}}^{0}$ in \ce{FeWO4} is a complex of $V_{\rm{Fe}}^{2-}$ (a defect created by the removal of a Fe$^{2+}$ ion) and two $\eta_{\rm Fe}^{+}$ with a binding energy of 0.79 eV; in \ce{MnWO4}, $V_{\rm{Mn}}^{0}$ is a complex of $V_{\rm{Mn}}^{2-}$ and two $\eta_{\rm Mn}^{+}$ with a binding energy of 0.84 eV; see Table~\ref{tab:formenergy}. In all cases, the identity of a defect (and its constituents) is determined via a detailed analysis of the induced charge density, magnetic moments, and lattice environment.

Energetically, we find that the lowest-energy defects in \ce{FeWO4} are $\eta_{\rm Fe}^{+}$ and $V_{\rm{Fe}}^{2-}$ [under conditions associated with points A and F in Fig.~\ref{fig;chempot}(a)] or $\eta_{\rm Fe}^{+}$ and $\eta_{\rm W}^{-}$ (points B$-$E). In \ce{MnWO4}, the lowest-energy defects are $\eta_{\rm Mn}^{+}$ and $V_{\rm{Mn}}^{2-}$ [points A, B, and F in Fig.~\ref{fig;chempot}(b)], $\eta_{\rm Mn}^{+}$ and $\eta_{\rm W}^{-}$ (point C), $V_{\rm O}^{2+}$ and $\eta_{\rm W}^{-}$ (point D), or W$_{\rm Mn}^{2+}$ and $\eta_{\rm W}^{-}$ (point E); see Table~\ref{tab:formenergy}. Under realistic conditions (see Sec.~\ref{subsec;chempot}), the dominant native defects in \ce{FeWO4} are thus $\eta_{\rm Fe}^{+}$ and $V_{\rm{Fe}}^{2-}$ with the formation energy of as low as 0.48 eV, whereas $\eta_{\rm Mn}^{+}$ and $V_{\rm{Mn}}^{2-}$ are the dominant defects in \ce{MnWO4} with the formation energy of as low as 0.74 eV. These oppositely charged defect pairs can occur in the form of $V_{\rm{Fe}}^{0}$ in \ce{FeWO4} or $V_{\rm{Mn}}^{0}$ in \ce{MnWO4}. With such low formation energies, the dominant defects will occur with high concentrations during preparation at high temperatures. They are expected to remain trapped in the materials even after cooling down to room temperature; the hole polarons can then act as preexisting charge-carrying defects during electrical conductivity measurements.

\subsection{Electronic conduction}

From the defect landscapes presented in Sec.~\ref{subsec;defect}, we find that certain native defects have non-negative formation energies only in a small range of the Fermi-level values near midgap; see, e.g., Figs.~\ref{fig;fe;fewo4} and \ref{fig;fe;mnwo4}. Like in other complex oxides,\cite{Hoang2011,Hoang2012274} \ce{FeWO4} (\ce{MnWO4}) thus cannot be p- or n-doped like a conventional semiconductor; any attempt to deliberately shift the Fermi level to the VBM or CBM via doping will lead to spontaneous formation of native defects that counteract the effects of shifting. Also, as evidenced from the results presented in Sec.~\ref{subsec;defect} a change from one (nominal) defect charge state to another is associated with polaron formation, indicating that the native defects cannot act as sources of free carriers. The electronic conduction in the tungstates is thus expected to occur via hopping of polarons.  

The migration of a small polaron between two positions $q_{\rm A}$ and $q_{\rm B}$ can be described by the transfer of its lattice distortion.\cite{Rosso2003} We estimate the migration barrier ($E_m$) by computing the energies of a set of supercell configurations linearly interpolated between $q_{\rm A}$ and $q_{\rm B}$ and identify the energy maximum. In \ce{FeWO4}, the migration barriers of $\eta_{\rm Fe}^+$ and $\eta_{\rm W}^{-}$ are found to be 0.14 and 0.12 eV, respectively; $E_m$ = 0.28 eV for $\eta_{\rm Mn}^+$ and 0.06 eV for $\eta_{\rm W}^{-}$ in \ce{MnWO4}. All these energy barriers are obtained for migration paths that go along the zigzag metal chain ($c$-axis in Fig.~\ref{fig;struct}). $\eta_{\rm W}^{-}$ thus has a lower migration barrier than $\eta_{\rm Fe}^+$ ($\eta_{\rm W}^{-}$), which is consistent with the fact that the self-trapping energy of the former is smaller than that of the latter.     

From the calculated formation energies and migration barriers, one can estimate the activation energy for conduction associated with polaron hopping. In general, the electronic or ionic conductivity can be defined as $\sigma$ = $q m c$, where $q$, $m$, and $c$ are the charge, mobility, and concentration of the current-carrying defects, respectively. Similar to what has been discussed in Ref.~\citenum{Hoang2014} in the context of battery materials, the concentration $c$ can include both thermally activated and athermal defects,
\begin{equation}\label{eq:concen2}
c=c_a+c_t=c_a + c_0 \mathrm{exp}\left(-\frac{E^f}{k_{\rm B}T}\right),
\end{equation}
where $c_a$ is the athermal concentration consisting of defects that may preexist in the material, $c_t$ is the concentration consisting of defects that are thermally activated during conductivity measurements at finite temperatures, $c_{0}$ is a prefactor, and $k_{\rm B}$ is Boltzmann's constant. The mobility of the defects can also be assumed to be thermally activated,
\begin{equation}\label{eq:mobility}
m=m_{0} \mathrm{exp}\left(-\frac{E_{m}}{k_{\rm B}T}\right),
\end{equation}
where $m_{0}$ is a prefactor and $E_{m}$ is the migration barrier. When the thermally activated defects are dominant, i.e., $c_t \gg c_a$, the observed temperature-dependence of the conductivity will show an intrinsic activation energy $E_{a}$ = $E^{f}$ + $E_{m}$, which includes both the formation energy and migration barrier. When the athermal defects are dominant, i.e., $c_a$ $\gg$ $c_t$, the activation energy will include only the migration barrier part, i.e., $E_{a}$ = $E_{m}$.\cite{Hoang2014} 

In principle, both hole and electron polarons can contribute to the electronic conductivity. However, since in the tungstates the hole polarons are the dominant (preexisting) electronic defects, the hole polaron hopping mechanism is expected to be dominant, which also explains the p-type conductivity observed in experiments.\cite{Sieber1982,Sieber1983,Schmidbauer1991,Bharati1982,Dissanayake1989} The measured activation energy for \ce{FeWO4} was reported to be in the range 0.15$-$0.32 eV,\cite{Sieber1982,Sieber1983,Schmidbauer1991} depending on how the sample was prepared. The lower limit is almost equal to the migration barrier (0.14 eV) of $\eta_{\rm Fe}^+$ obtained in our calculations. In this case, it is likely that there is a high concentration of preexisting $\eta_{\rm Fe}^+$ in the sample and the activation energy contains only the $E_m$ term as discussed above. As for the other samples, the measured effective activation energy is larger than $E_m$ and expected to depend on the $c_a/c_t$ ratio. Schmidbauer et al.\cite{Schmidbauer1991} also observed lower $E_a$ values in samples prepared at higher oxygen partial pressures, consistent with our calculations showing that $E^f(\eta_{\rm Fe}^+)$ is lower at higher oxygen chemical potential values. In \ce{MnWO4}, the measured activation energy is in the range 0.53$-$0.65 eV, larger than $E_m(\eta_{\rm Mn}^+)$ as expected. Since $E^f(\eta_{\rm Fe}^+)$ $\lesssim$ $E^f(\eta_{\rm Mn}^+)$ (see Table~\ref{tab:formenergy}), the polaron $\eta_{\rm Fe}^+$ is expected to be more abundant in \ce{FeWO4} than $\eta_{\rm Mn}^+$ in \ce{MnWO4}, which also explains why the measured activation energy is lower in the former.           

\subsection{Charge-storage mechanism}

Given the materials' ability to form small polarons associated with the transition-metal ions in the bulk, we speculate that the polarons and the associated redox reactions can also occur at/near the surface. In fact, the formation energy of the polarons may even be lower at the surface than in the bulk, given the less constrained lattice environment. It is known that a pseudocapacitive charge-storage mechanism is characterized by the ability of the material to have reversible redox reactions at/near the surface when in contact with an electrolyte.\cite{Augustyn2014EES} Our results for the polarons in \ce{FeWO4} and \ce{MnWO4} appear to be consistent with the fact that the materials have successfully been used as electrodes in supercapacitors,\cite{Bretesche2015,Li2015} with the Fe$^{3+}$/Fe$^{2+}$ redox center specifically being suggested to be responsible for the electrochemical performance observed in \ce{FeWO4}.\cite{Bretesche2015} Further studies are, however, needed to develop a more detailed understanding of the occurrence of redox pseudocapacitance in the metal tungstates. It would be interesting to explore if the W ion can also be utilized for surface redox reactions.

\section{Conclusions}

We have carried out a hybrid density-functional study of bulk properties and defect physics in \ce{FeWO4} and \ce{MnWO4}. The lattice parameters, static dielectric constants, and band gaps obtained in our calculations are in good agreement with experiments. The electronic structure at the valence-band top is found to be predominantly the highly localized Fe (Mn) $3d$ states, whereas at the conduction-band bottom it is predominantly the localized W $5d$ states. These features of the electronic structure allow for the formation of both hole and electron polarons in the tungstates. Due to the presence of other negatively charged defects that have lower formation energies than the electron polarons, however, the dominant point defects in \ce{FeWO4} (\ce{MnWO4}) under realistic synthesis conditions are small hole polarons and negatively charged iron (manganese) vacancies. Electronic conduction thus occurs via hopping of the low-formation-energy and highly mobile hole polarons, consistent with the good p-type conductivity observed in experiments. In light of the results, we also briefly comment on the redox pseudocapacitance reported to occur in the tungstate materials.

\begin{acknowledgments}

The author is grateful to \'{E}cole Polytechnique de l'Universit\'{e} de Nantes (Polytech Nantes) for supporting his visit to Institut des Mat\'{e}riaux Jean Rouxel (IMN) during which these tungstate materials were brought to his attention and to Sylvian Cadars for discussions. This work was supported in part by the U.S.~Department of Energy Grant No.~DE-SC0001717 and made use of computing resources at North Dakota State University's Center for Computationally Assisted Science and Technology and the National Energy Research Scientific Computing Center (NERSC), a DOE Office of Science User Facility supported by the Office of Science of the U.S.~Department of Energy under Contract No.~DE-AC02-05CH11231.

\end{acknowledgments}


%

\end{document}